\documentclass{icrc29}
\usepackage{graphicx,amssymb,amsmath,times}
\newcommand{\Xmax}{X_{\max}}
\begin{document}
\title[Air Shower Properties with the Gluon Saturation Model BBL]{Air Shower Properties with the Gluon Saturation Model BBL}
\author[H.J. Drescher]{H.J. Drescher\\
        Frankfurt Institute for Advanced Studies (FIAS),
	Johann Wolfgang Goethe-Universit\"at,\\
        Max-von-Laue-Stra{\ss}e 1, 60438 Frankfurt am Main
        }
\presenter{Presenter: H.J. Drescher (h.j.drescher@fias.uni-frankfurt.de), HE.1.4}

\maketitle

\begin{abstract}
The hadronic interaction model BBL implements the ideas of gluon
saturation due to large densities. When approaching the black body
limit at high energies, leading partons acquire large transverse
momenta which breaks up their coherence. This leads to a suppression
of forward scattering, and is therefore important for air showers. We
discuss some general aspects of this new approach and their influence
on air shower properties as seen by fluorescence and surface
detectors: The position of the shower maximum is reduced due to 
stronger absorption in the atmosphere. The lateral distribution
functions become flatter for the same reason. Muons are produced
abundantly due to high multiplicities in the mid-rapidity region. The
response of water Cherenkov detectors and comparisons to other
interaction models are shown.
\end{abstract}

\section{Introduction}

In this paper, we discuss the high energy limit of hadron nucleus
scattering and their influence on air shower properties, as described
in Ref. \cite{Drescher:2004sd}. When
approaching highest energies, we expect the differential scattering
amplitude to become close to unity. The partons acquire a large
transverse momentum of the order of the saturation scale, which leads to an
steeper spectrum of forward scattered particles, the most
important phase space region for air shower properties.

All air showers in this paper have been computed with the Seneca model
\cite{seneca} using UrQMD~1.3.1 \cite{urqmd} as low-energy model below 100
GeV. 

\section{High energy hadron nucleus scattering}

\begin{figure}[b]
\begin{center}
\includegraphics*[width=0.49\textwidth,angle=0,clip]{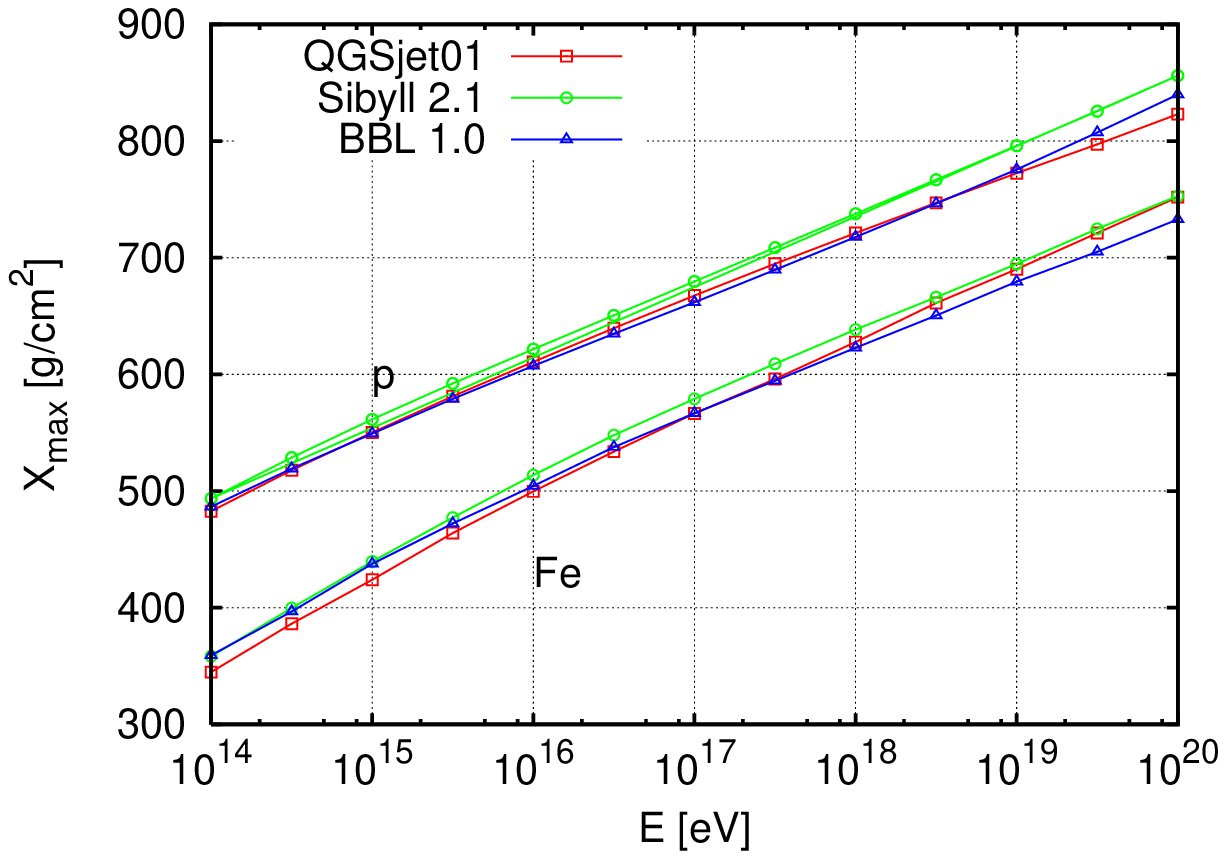}
\includegraphics*[width=0.49\textwidth,angle=0,clip]{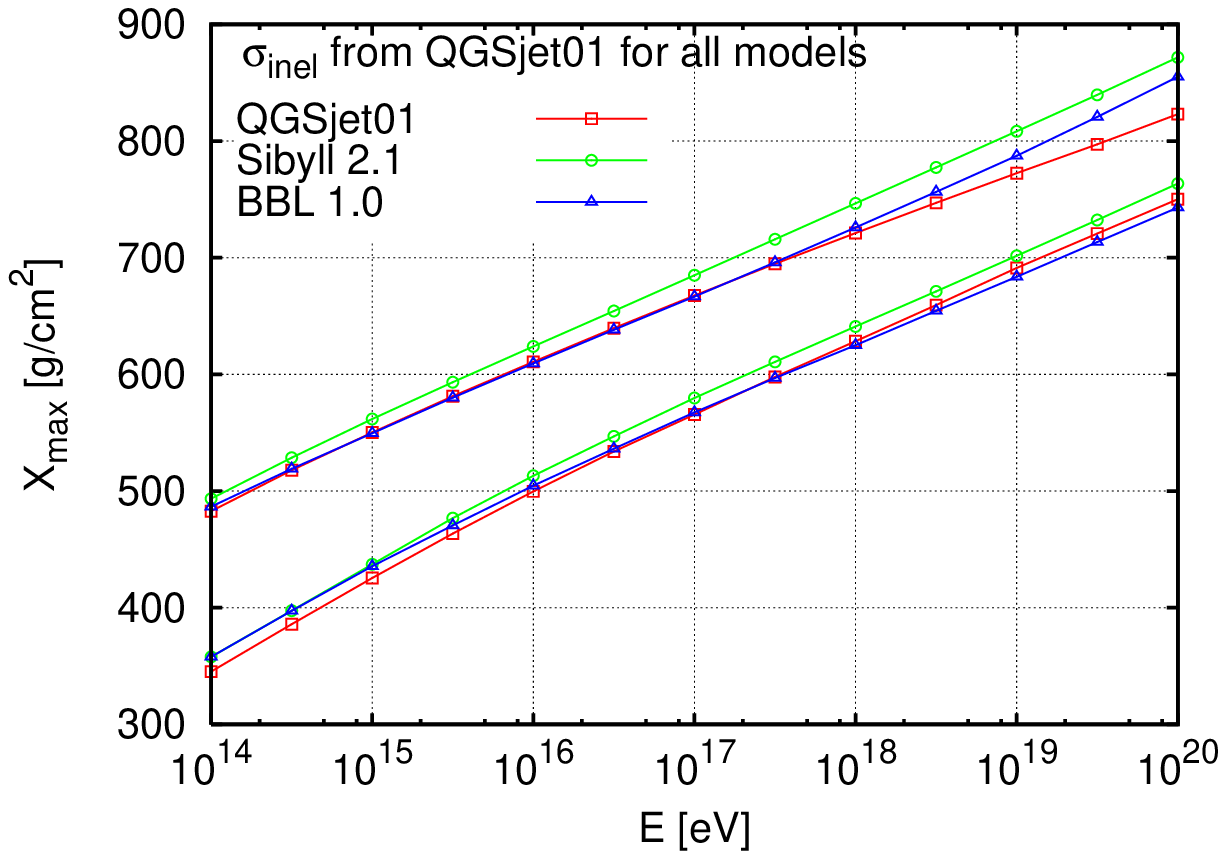}
\caption{\label {fig2} Left panel: $\Xmax$ as a function of
  energy. Right panel: $\Xmax$ of models using inelastic
  cross-sections from QGSjet01, i.e. differences are due to forward
  scattering only.}
\end{center}
\end{figure}

When approaching highest energies, higher twist corrections in hadron
nucleus collisions become increasingly important. Attempts to account
for this are the implementation of an energy-dependent $p_t$ cutoff
\cite{Sibyll} for hard scattering or the resummation of enhanced
pomeron diagrams in an efficient manner \cite{Ostapchenko:2004ss}.

Our approach is to consider the black disk limit (or black body limit
- BBL) at high gluon densities within the Color Glass Condensate (CGC)
approach\cite{sat}, where the interaction probability is close to
unity. Within this model, the scattering amplitude of a quark on a
dense gluon field is resummed to all orders. This can be done since
the typical scale of the gluon density, the so-called saturation
momentum $Q_s$ is larger than $\Lambda_{\rm QCD}$ and weak coupling
methods are applicable. 

The Monte Carlo interaction model, which incorporates the physics
discussed here has been introduced in Ref.\cite{Drescher:2004sd} and
discussed in more detail in Ref. \cite{Drescher:2005ak}. The most
important feature is the suppression of forward scattering. As a
consequence of energy conservation we also find an increase of
particle production at lower rapidities. Although less important for
the longitudinal profile of an air shower, this results in an
increased muon production and is therefore important to be considered
when deducing composition from this observable.

Since this approach is only valid when the saturation momentum is
large, we use this model only in the high energy/high density
limit. For low energies and peripheral interactions we use the
standard pQCD Monte Carlo model Sibyll \cite{Sibyll}.

\section{Longitudinal Profile}

\begin{figure}[t]
\begin{center}
\includegraphics*[width=0.49\textwidth,angle=0,clip]{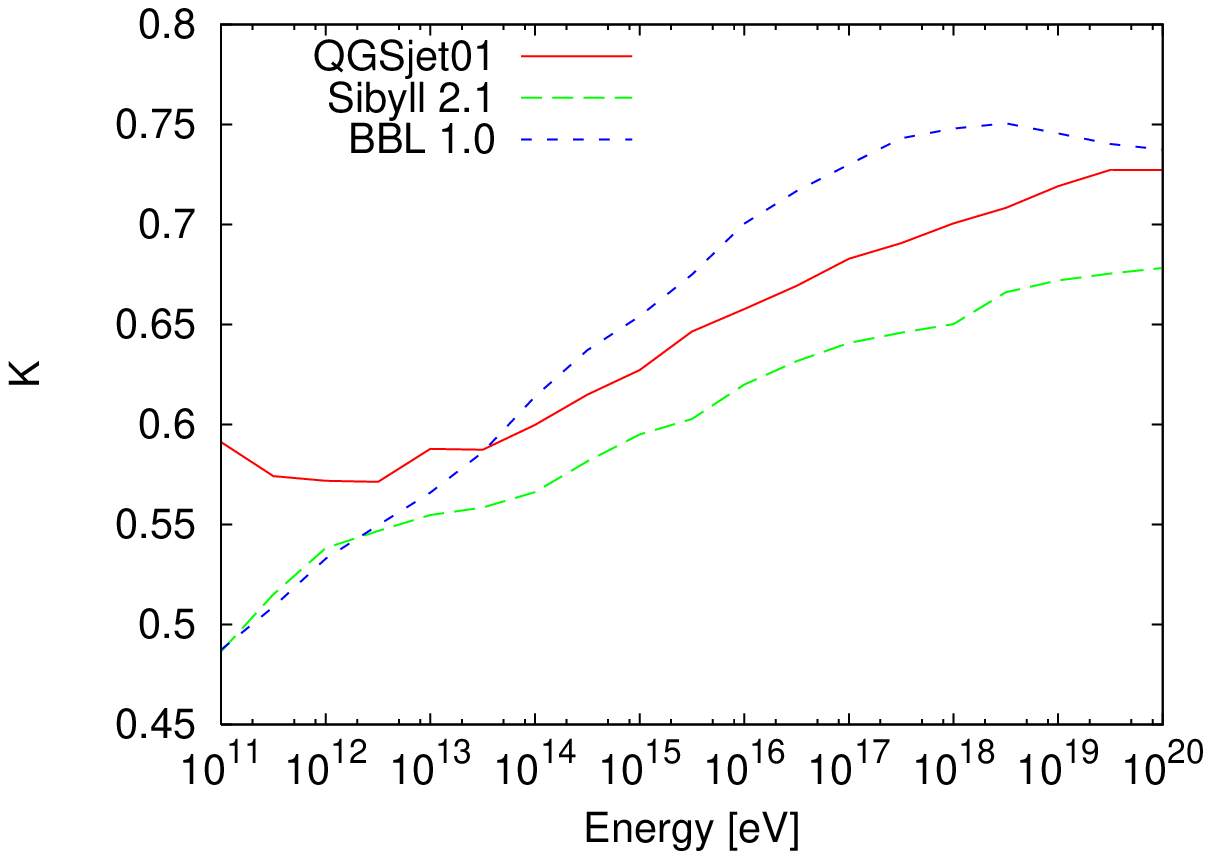}
\includegraphics*[width=0.49\textwidth,angle=0,clip]{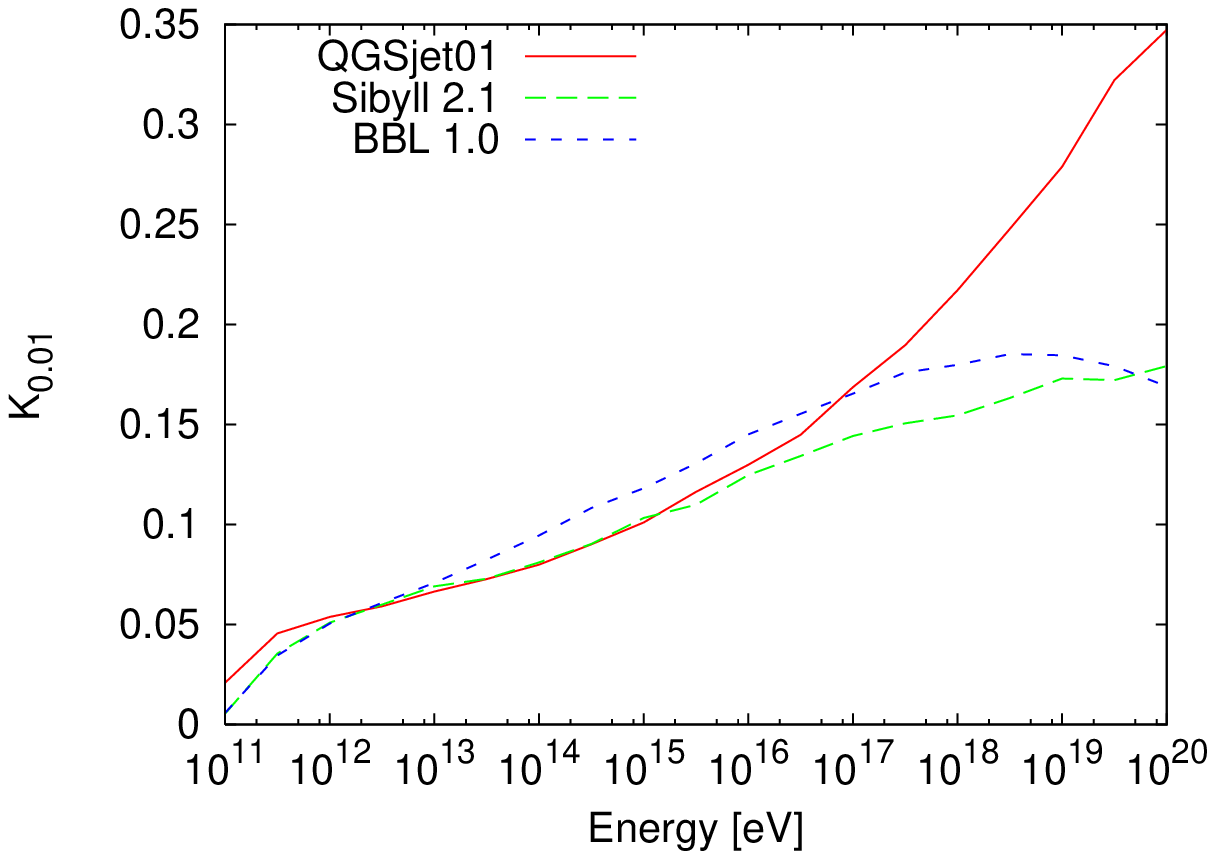}
\caption{\label {fig1} Left panel: Inelasticity of the three models as a function
  of primary energy. Right panel: Inelasticity $K_{0.01}$, evaluated
  in the forward phase space region $0.01 \le x_F \le 1 $.}
\end{center}
\end{figure}

Results for the mean $\Xmax$ are shown in Fig. \ref{fig2}.
The longitudinal profile is dominated by mainly two observables, the
forward scattering, often described by the inelasticity ($K=1-x_F$ of
the fastest particle) and the inelastic cross section which determines the
mean free path. Since we have used the inelastic cross
sections from Sibyll, differences to this model are due to the reduced forward
scattering solely, whereas the difference to QGSjet01 \cite{qgsjet}
comes from a combination of both different cross sections and
inelasticity.

Also shown are the results for iron induced showers
computed with the BBL model by simple superposition of nucleon-air
collisions, just as in Sibyll. This should give a good
estimate, since the the running coupling evolution is less sensitive
to initial conditions, i.e. the density of the projectile. The full
implementation of nucleus-nucleus collisions with consideration of
projectile saturation scale will be left for the future. 

Fig. \ref{fig1} shows the inelasticity of the models, defined as 
mean $1-x_F$ of the fastest particle. The right panel
shows $K_{0.01} = 1-\int_{0.01}^1 x_F dn/dx_F dx_F$, which corresponds
to the inelasticity averaged over a larger region in forward phase
space, not only the fastest particle. 

The right panel in Fig. \ref{fig2} shows $\Xmax$ for Sibyll and BBL
when one applies inelastic cross-sections from QGSjet01 which are
somewhat smaller, see e.g. \cite{Engel:2005gf}. At largest energies
the $\Xmax$ of Sibyll and BBL get shifted by ca. 15~g/cm$^2$ and reach 50 and
30~g/cm$^2$ larger depth than QGSjet, correspondingly.  Assuming the same
inelastic cross-sections, BBL has a larger $\Xmax$ than
QGSjet01, despite a larger inelasticity of the former. But as can be seen in
Fig. \ref{fig1}, QGSjet01 has a larger $K_{0.01}$, as a consequence of
a rather flat $x$-distribution at these energies. Clearly, the
inelasticity alone is not sufficient to characterize forward
scattering. In Ref. \cite{Drescher:2005ak} is was shown that $\Xmax$ is
sensitve to the forward region down to $x_F \approx 10^{-2} -
10^{-3}$. Therefore the $K_{0.01}$, or even $K_{0.001}$ variable is
more suitable to describe this specific range. 

\section{Muon- and electron numbers}

\begin{figure}[t]
\begin{center}
\includegraphics*[width=0.49\textwidth,angle=0,clip]{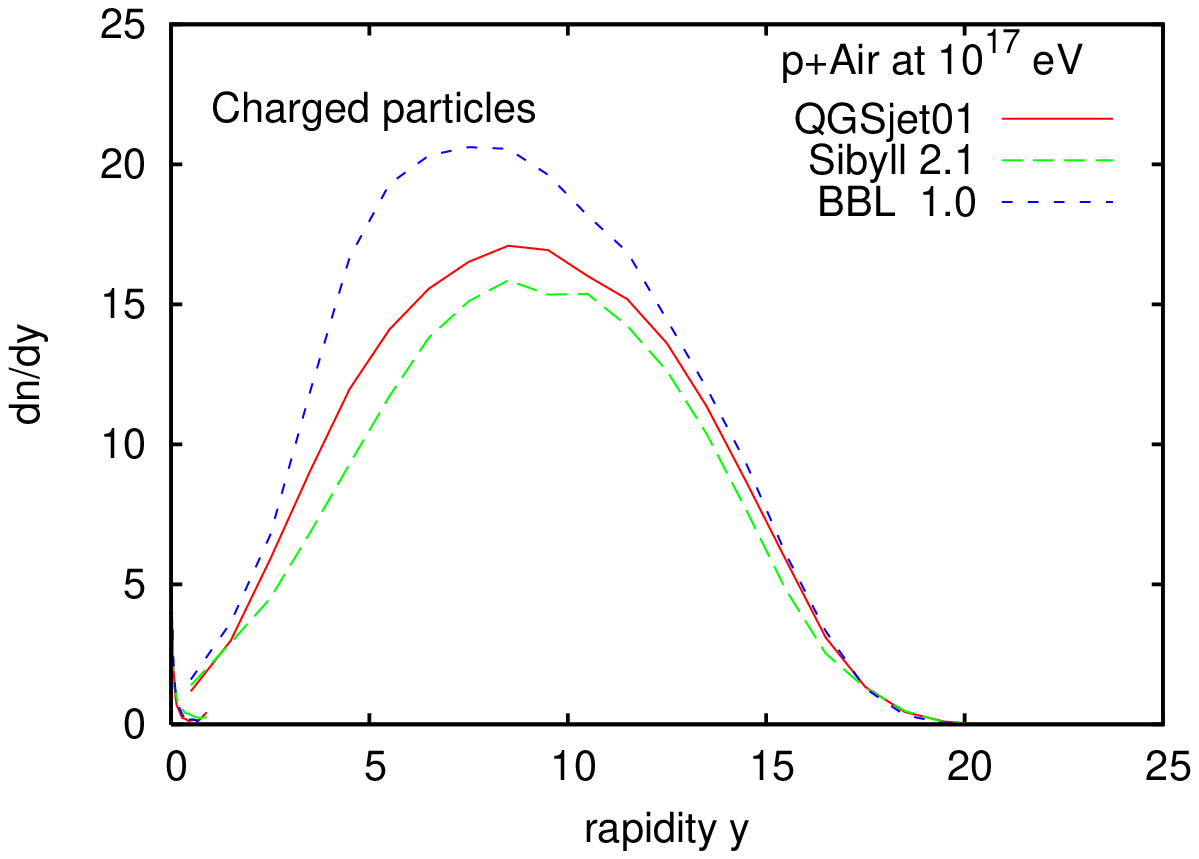}
\includegraphics*[width=0.49\textwidth,angle=0,clip]{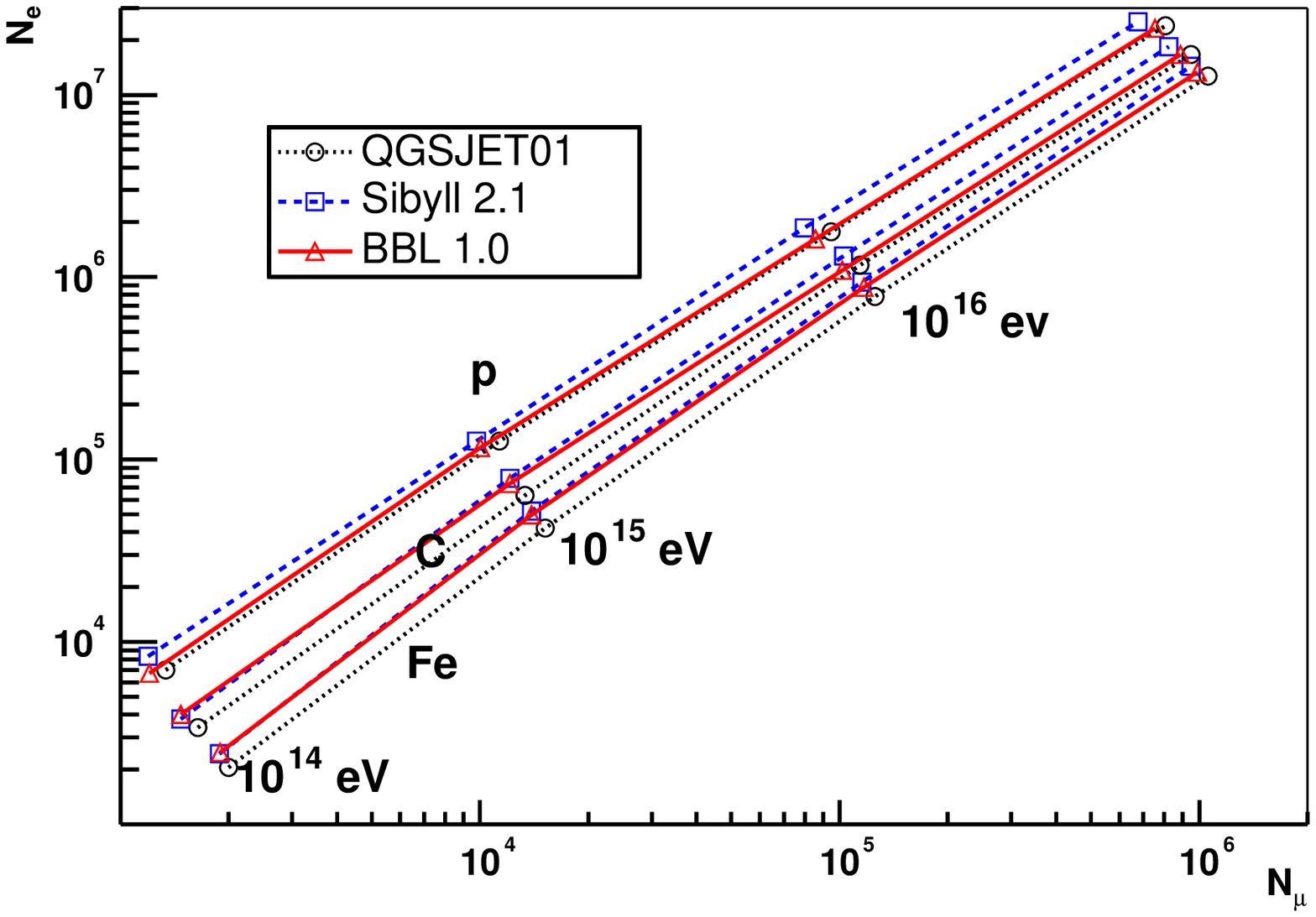}
\caption{\label {fig:nmne} Left: Rapidity of charged particles for
  p+Air at $10^{17}$ eV. Right: Electron/positron and muon numbers in the
  intermediate energy range.}
\end{center}
\end{figure}

The major consequence of the implementation of the black body limit is
the suppression of forward scattering. However, this also leads to an
enhancement of the multiplicity in the mid-rapidity region, see
Fig. \ref{fig:nmne}. This has almost no influence on the longitudinal
profile, but increases the muon numbers as decay products of
low-energy particles.  A plot of mean electron/positron and muon
numbers is shown in the right panel of Fig. \ref{fig:nmne}. The BBL
model seems to interpolate somewhat between the results of Sibyll and
QGSjet01, with the transition happening at about $10^{14}$~eV. Since the
primary energy is shared between the nucleons of the projectile
nucleus, this interpolation also happens as a function of mass number
$A$ at correspondingly higher energy. This can be seen for the carbon
projectile in the knee region.  Whether this model can help to resolve
known problems of models in KASCADE, is subject to a detailed
comparison.

\section{Lateral Distribution Function}

\begin{figure}[t]
\begin{center}
\includegraphics*[width=0.49\textwidth,angle=0,clip]{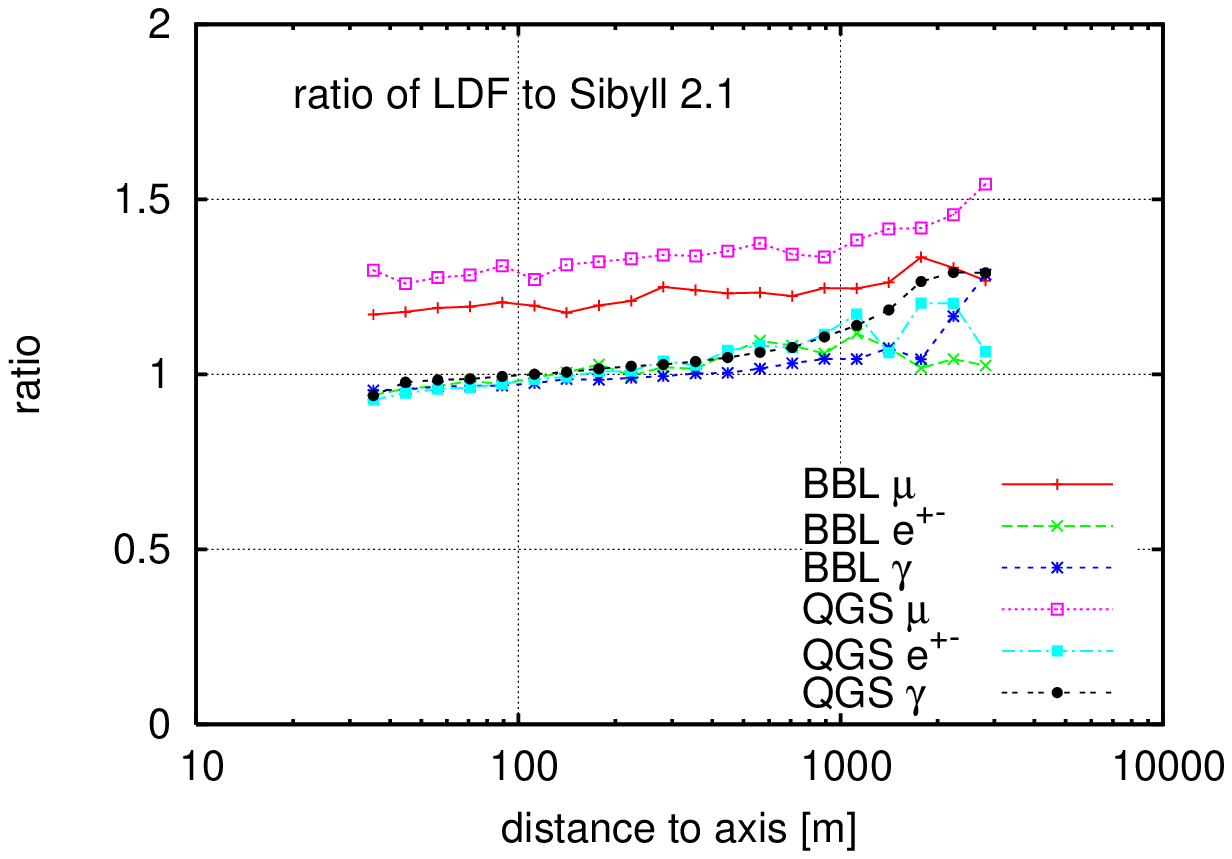}
\includegraphics*[width=0.49\textwidth,angle=0,clip]{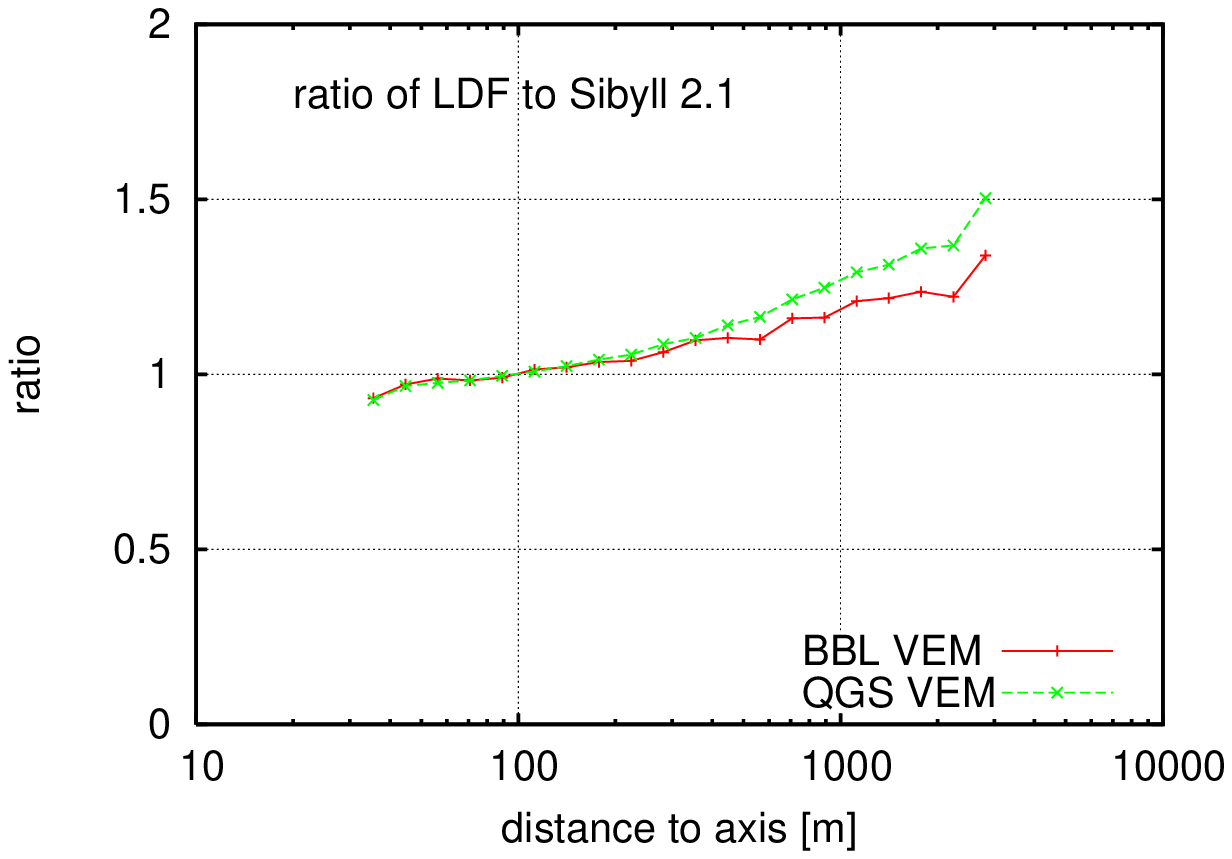}
\caption{\label {fig:lat} Lateral distribution functions of different
  particles and of the total signal in VEM units. Results are shown 
  as a ratio to the LDFs from Sibyll.
  The low energy model ($E<100$ GeV) is UrQMD.}
\end{center}
\end{figure}

In Fig. \ref{fig:lat} we show the lateral distribution
functions for $10^{19}$~eV vertical proton showers at Auger
altitude. Ratios to the LDFs of Sibyll are plotted. 
Besides from a small decrease of the slope for electrons and photons,
most noticeable is the increased muon number. QGSjet01 shows the same
qualitative behavior as BBL. Since muons dominate the VEM signal at
large distances, the combined LDF in VEM units shows a significantly
flatter slope. 

\subsection*{Acknowledgements}

The author thanks Adrian Dumitru and Mark Strikman for useful
discussions. This work has been supported by BMBF grant DESY~05CT2RFA/7.

\end{document}